 \documentclass[authoryear,preprint,12pt]{elsarticle}

 \usepackage{graphicx}

\usepackage{amssymb}

\usepackage{lineno}

\journal{Planetary and Space Science}

\begin{document}

\begin{frontmatter}





\title{Small crater populations on Vesta}

%
\author[mar,kri]{S. Marchi\corref{cor1}}
\ead{marchi@boulder.swri.edu}
\author[mar]{W.F. Bottke}
\author[obr]{D.P. O'Brien}
\author[sch]{P. Schenk} 
\author[mot]{S. Mottola}    
\author[cri]{M.C. De Sanctis} 
\author[kri]{D.A. Kring}
\author[wil]{D.A. Williams}
\author[ray]{C.A. Raymond} 
\author[rus]{C.T. Russell} 

%
%
\cortext[cor1]{Corresponding author}

\address[mar]{NASA Lunar Science Institute, Southwest Research Institute, Boulder, CO, USA}
\address[obr]{Planetary Science Institute, Tucson, AZ, USA}
\address[sch]{Lunar and Planetary Institute, Houston, TX, USA}
\address[mot]{Institut f\"ur Planetenforschung, DLR-Berlin, Germany}
\address[cri]{Istituto Nazionale d'Astrofisica, Roma, Italy}
\address[kri]{NASA Lunar Science Institute, USRA - Lunar and Planetary Institute, Houston, TX, USA}
\address[wil]{Arizona State University, Tempe, AZ, USA}
\address[ray]{Jet Propulsion Laboratory, California Institute of Technology, Pasadena, CA, USA}
\address[rus]{University of California, Los Angeles, CA, USA}

\begin{abstract}

The NASA Dawn mission has extensively examined the surface of asteroid
Vesta, the second most massive body in the main belt. The high quality
of the gathered data provides us with an unique opportunity to
determine the surface and internal properties of one of the most
important and intriguing main belt asteroids (MBAs). In this paper, we
focus on the size frequency distributions (SFDs) of sub-kilometer
impact craters observed at high spatial resolution on several {
  selected young} terrains on Vesta. These small crater populations
offer an excellent opportunity to determine the nature of their
asteroidal precursors (namely MBAs) at sizes that are not directly
observable from ground-based telescopes (i.e., below $\sim$100~m
diameter). Moreover, unlike many other MBA surfaces observed by
spacecraft thus far, the { young} terrains examined had crater
spatial densities that were far from empirical saturation. Overall, we
find that the cumulative power-law index (slope) of small crater SFDs
on Vesta is fairly consistent with predictions derived from current
collisional and dynamical models down to a projectile size of
$\sim$10~m diameter \citep[e.g.][]{bot05a,bot05b}.  The shape of the
{ impactor SFD for small projectile sizes} does not appear to have
changed over the last several billions of years, and an argument can
be made that the absolute number of small MBAs has remained roughly
constant (within a factor of 2) over the same time period.  The
apparent steady state nature of the main belt population potentially
provides us with a set of intriguing constraints that can be used to
glean insights into the physical evolution of individual { MBAs} as
well as the { main belt} as an ensemble.

\end{abstract}


\end{frontmatter}



\section{Introduction}

The NASA Dawn { spacecraft} was conceived to address key questions
related to the {\it dawn} of our solar system, hence the name of the
mission. The spacecraft was launched on the 27 September 2007.  Its
main destinations were asteroids Vesta and Ceres, the two most massive
bodies in the main belt of asteroids \citep{bri02}.

Vesta and Ceres were chosen because they are excellent targets for
studying early solar system processes.  They have also bore witness to
4.5~Gyr of main belt evolution, with cratered terrains that
allow us to glean insights into how processes like collisional
evolution have shaped the asteroid belt. In this paper, we focus our
attention on this aspect of main belt history, and in particular on
the size-frequency distribution of small main belt asteroids (MBAs).

Small MBAs, defined here as bodies with diameters less than a few
kilometers, are thought to be a steady-state population of fragments
made by collisional and dynamical evolution processes
\citep[e.g.][]{doh69,dav94,obr05,bot05a,bot05b}.  The main source of
small MBAs { are asteroid} impacts, though a sizable fraction may
come from mass shedding events caused by the YORP effect, a
non-gravitational force that describes how the re-radiation of
sunlight can spin some asteroids up to the fission limit
\citep[e.g.][]{pra10}.  The sinks for small MBAs are collisions, YORP,
and thermal Yarkovsky drift forces, which drive many small MBAs to
dynamical resonances where they can escape the main belt and reach
planet-crossing orbits \citep[e.g.][]{bot06}. These mechanisms work
incessantly to resupply the projectile populations that have struck
the Moon and terrestrial planets over the last several Gyr.

By understanding the MBAs size distribution and how it has
varied over time, we can glean insights into main belt history, the
physics of asteroid fragmentation, the importance of YORP-driven mass
shedding events, and how much the near-Earth asteroid (NEA) population
has changed over billions of years.  A problem with this, however, is
that the current and past nature of the MBA population is only
partially (or indirectly) constrained.

For example, ground-based surveys have so far only been able to
determine the MBA population down to objects a few hundreds of meters
in diameter \citep[e.g.][]{ive01,yos07,gla09}. { To deduce the
  nature of the MBA population at small sizes, collisional and
  dynamical models have tried to use a wide array of constraints, like
  the number and nature of the observed main belt asteroids and
  asteroid families, the known near-Earth asteroid population, small
  impact events detected in the Earth's upper atmosphere and the
  cratering history of the Moon and terrestrial planets
  \citep[e.g.][]{mor03,bot05a,bot05b,obr05}}. This has led to many
intriguing solutions, but there has been no way to verify their work.

Ideally, a more direct way to infer the MBA population and how it has
changed is to examine the crater populations on the surfaces of main
belt asteroids. { A number} of asteroids have been visited by
spacecraft (see Table~\ref{tab1}). While this has led to an enormous
reservoir of information on the collisional history of individual
asteroids, interpreting the cratering record of these bodies is often
difficult \citep[e.g.][]{obr06}.  Many of the larger worlds (e.g. Ida,
Mathilde, Lutetia) have ancient surfaces, such that the spatial
density of craters on their surfaces are arguably close to or perhaps
in saturation, defined as a state where newly-formed craters
obliterate pre-existing ones
\citep[][]{gau70,har84,cha86,ric09,mar12a}. Craters on smaller
asteroids, like Eros, Gaspra, or Steins, have possibly been affected
by processes like impact-induced seismic shaking that can erase small
craters \citep[e.g.][]{gre94,ric04,ric05,obr06}.  For these and other
reasons, previously visited asteroids, at first glance, do not provide
a clear and consistent record of the MBA population.

Vesta offers a unique opportunity to { explore the small MBA
  population as a function of time.}  Vesta offers several advantages
over the previously mentioned asteroids explored by spacecraft.
First, it is large enough that only the largest, closest impacts to
{ the counting area} are likely to produce episodes of extensive
seismic shaking, { unlike smaller asteroids where impacts anywhere
  on the surface may have global effects}. Second, Vesta's large
surface area allows for the existence of adjacent regions with very
different ages, ranging from heavily cratered units in the northern
hemisphere to mildly cratered units that mainly exist in the southern
hemisphere. The latter regions appear to have been locally reset by
relatively { recent large impact events}. Accordingly, it is
possible to study populations of craters made by small MBAs well
before they can reach saturation.

\begin{table*}[H!]
\caption{Asteroids visited by spacecraft in the past. The column
  ``Comment'' contains a brief summary concerning the observations of
  sub-kilometer craters, which is the focus of this paper. The
  comments summarize results from the following papers:
  \cite{mar10,mar12c} (Steins, Lutetia); \cite{cha96a,cha96b,cha99}
  (Gaspra, Ida, Mathilde); \cite{ric04} (Eros).  The symbol $\dag$
  marks near-Earth objects.}
\label{tab1}
\begin{center}
\def\V{\rule{0pt}{3.5ex}}
\def\P{\rule{0pt}{2.3ex}}
\begin{tabular}{c|lll} 
\hline \hline 
\V Asteroid & \multicolumn{1}{c}{Average size (km)} & \multicolumn{1}{c}{Spacecraft} & \multicolumn{1}{c}{Comment} \\[1.3
ex]
\hline
Lutetia       & 100  &   Rosetta  & Some data available \\
Mathilde      & 50   &   NEAR     & Saturated, poor resolution \\
Eros$^\dag$    & 17   &   NEAR     & Small craters erased \\
Ida           & 16   &   Galileo  & Saturated \\
Gaspra        & 6    &   Galileo  & Some data available \\
Steins        & 5    &   Rosetta  & Poor resolution, small craters erased \\
Itokawa$^\dag$ & 0.3  &   Hayabusa & Too small object\\[1.3ex]
\hline \hline 
\end{tabular}
\end{center}
\end{table*}

\section{Vesta's Sub-Kilometer Crater Populations}

The Dawn spacecraft imaged the surface of Vesta at varying spatial
resolutions during the orbiting phase of its mission. In this work, we
concentrate on the Low Altitude Mapping Orbit (LAMO) phase which
lasted for 141 days, from 12 December 2011 to 30 April 2012. The
spacecraft operated at an altitude of about 210~km, resulting in an
average { Framing Camera} spatial resolution of approximately
20~m/px. At this scale, the surface of Vesta appears peppered by
numerous sub-kilometer craters, though only some are fresh; the rest
appear to be partially buried by regolith. Most of these terrains are
poorly suited to deduce the small MBA population, with many close to
or in saturation.  In addition, many craters on these surfaces have
been erased, possibly as a consequence of redistributed ejecta and/or
seismic shaking episodes via large impacts.

\begin{figure*}[h]
\includegraphics[width=9cm]{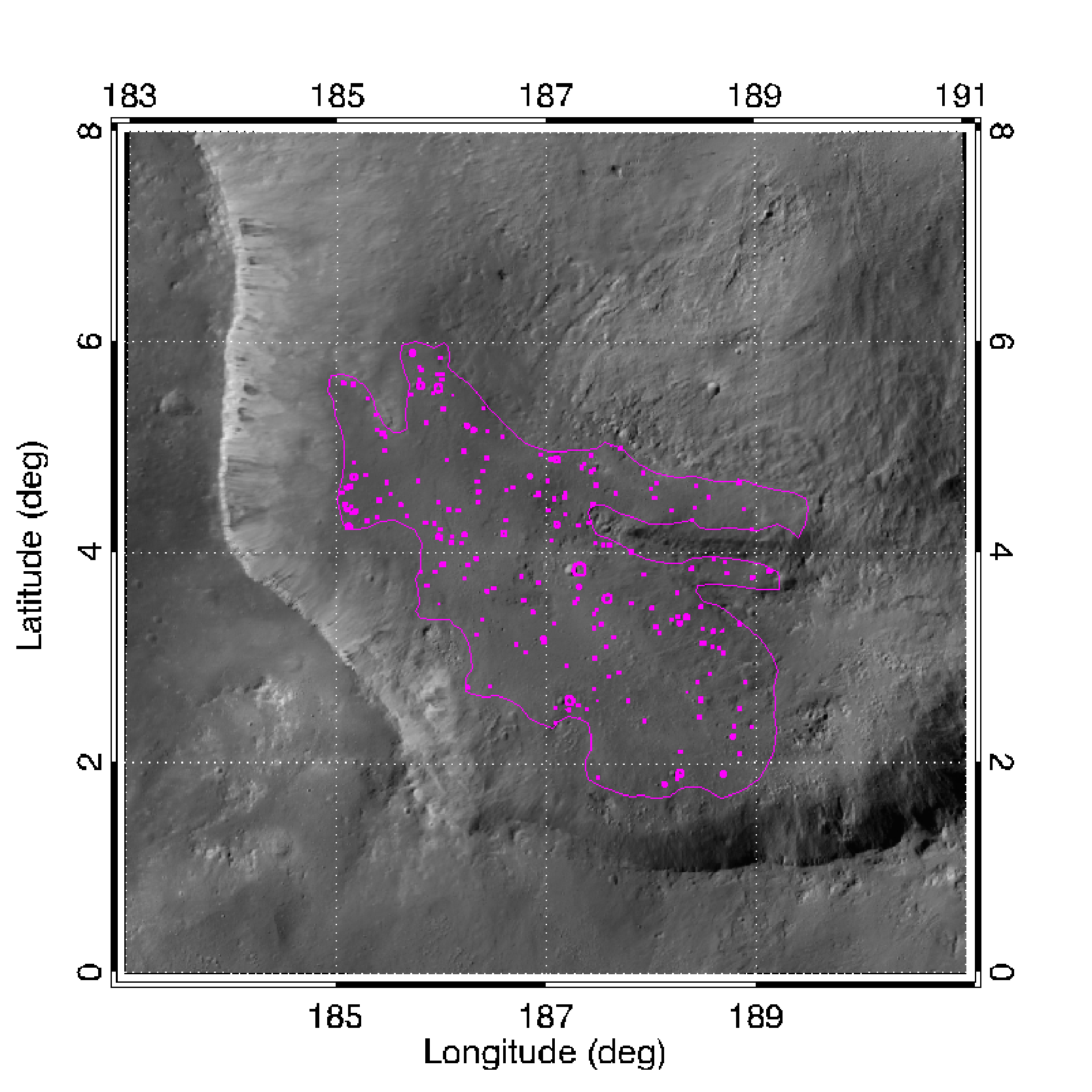}
\caption{Orthographic projection of a { LAMO mosaic of the
    southwest} portion of the 60-km Marcia crater.  The contour line
  marks the region used for crater counts, whose area is 194.5~km$^2$.
  The map has a resolution of $\sim$15~m/px. Circles indicate the 206
  craters that have been measured, ranging from 50~m to 500~m in
  diameter.}
\label{marcia_lamo}
\end{figure*}

To overcome these concerns, we turn our attention to two of the
youngest terrains on Vesta that appear to be relatively undisturbed by
post-emplacement evolution.  The first one is associated with the
fresh 60-km diameter Marcia crater \citep{wil12}.  Located
near Vesta's equator, Marcia crater is characterized by the presence
of pitted terrains in the proximal ejecta and crater floor possibly
made by the outgassing of volatile-rich material \citep{den12}.
Marcia's ejecta blanket is very ragged as a result of mantling and
partial erosion of previous topography. These characteristics make it
difficult to detect sub-kilometer craters on many { Marcia-related}
units.  To avoid these problems, we focused on a relatively small and
smooth unit -possibly impact melt- within the rim of the Marcia crater
shown in Fig.~\ref{marcia_lamo} \citep{wil12}.  Our crater
counts are shown on the figure. In defining the crater counting unit,
we were careful to stay away from high-slope terrains (e.g., crater
walls) { because mass movements may affect the preservation of
  small craters.}

\begin{figure*}[h]
\includegraphics[width=15cm]{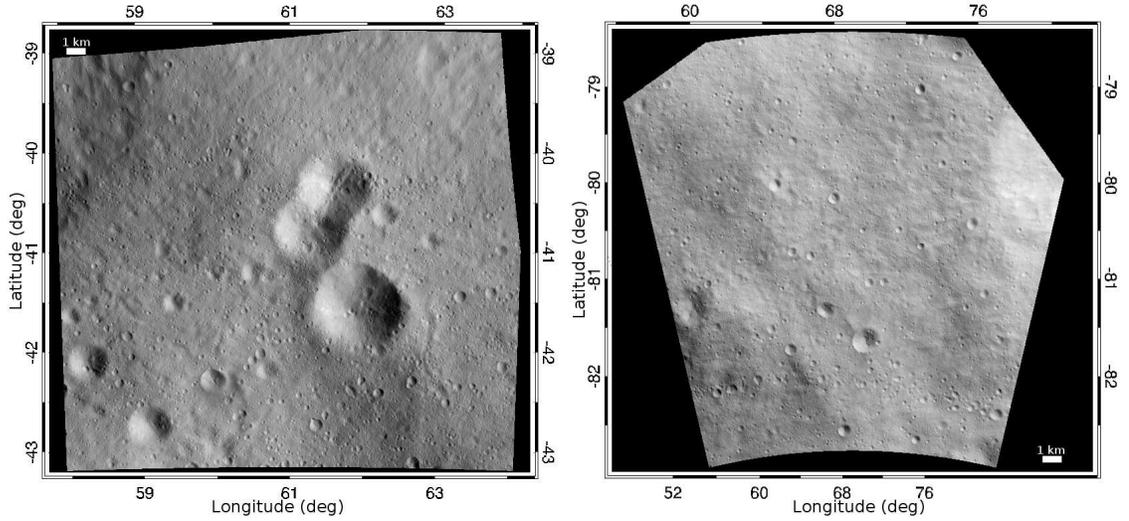}
\caption{Orthographic projections of two high resolution images
  ($\sim$22~m/px) acquired during the LAMO phase. Image ID:
  FC21B0014580\_11353023506F1C (left panel: { ejecta})
  FC21B0015216\_11359132403F1D (right panel: { floor}). }
\label{rs_floor_ejecta}
\end{figure*}

The second terrain is associated with the Rheasilvia basin, a 500-km
diameter impact basin that dominates the southern hemisphere of Vesta
\citep[e.g.][]{sch12}.  A preliminary analysis, based on the
superposed crater SFD on its floor, showed that Rheasilvia's age is
approximately 1~Gyr \citep{mar12b}. Figure~\ref{rs_floor_ejecta} shows
two LAMO images from Rheasilvia's floor and its proximal ejecta
blanket. A comparison between the images shows that both terrains, at
a first sight, appear to be well suited for counting fresh
small craters. The Rheasilvia floor, however, is characterized by
high slopes, while a large portion of the ejecta blanket near the
rim known as Matronalia Rupes\footnote{See IAU link
  http://planetarynames.wr.usgs.gov/Page/VESTA/target for a list of
  official names of Vestan surface features.} is remarkably smooth and
flat.  This indicates that the emplacement of the ejected material
created a layer that may be kilometers deep, thick enough to
obliterate pre-existing topography \citep{sch12}.  For these
reasons, we opted to measure small craters only on the ejecta blanket.

\begin{figure*}[h]
\includegraphics[width=7cm]{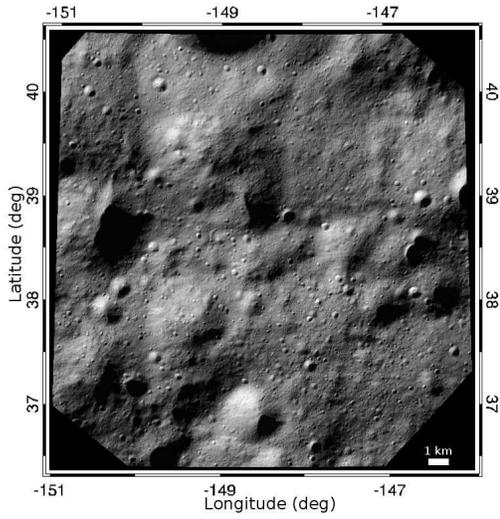}
\caption{Orthographic projection of a LAMO image acquired in the
  northern hemisphere ($\sim21$~m/px). Image ID:
  FC21B0023439\_12085231933F1A.}
\label{nh}
\end{figure*}

For comparison, Figure~\ref{nh} shows a representative LAMO image
acquired in the heavily cratered northern hemisphere \citep{mar12b}
that has a comparable resolution to those acquired in the southern
hemisphere. Compared to the relatively fresh terrains shown in
Figure~\ref{rs_floor_ejecta}, here we see a highly complex topography,
with numerous multi-kilometer depressions that could be ancient impact
craters largely obliterated by subsequent evolution. Many of the
sub-kilometer craters also look subdued or partially buried. In these
terrains, we find it plausible that billions of years of cratering
events have led to a quasi-steady state between crater formation and
removal. The crater spatial densities determined here will be
challenging to decipher, and we save them for future work.
 
Figure~\ref{rs_lamo} shows all of the craters that we have mapped on
the smooth Rheasilvia ejecta unit. We concentrated our attention on a
particular sub-region that was large enough (and contained enough
craters) to be representative of the entire ejecta unit.

\section{The Model Production Function Chronology}

The crater size-frequency distributions (SFDs) of the units above can
be compared to model size distributions of the MBAs in order to derive
their crater retention ages.  Here we use the chronology framework
provided by the Model Production Function (MPF)
\citep{mar09,mar10,mar11}. By modeling the MBA impactor flux and
transforming the results using a crater scaling relationship, we can
compare our crater production function to the data and solve for the
surface age.  This method is valid provided our MPF accurately
estimates the MBA population over time.
  
\begin{figure*}[h]
\includegraphics[width=12cm,angle=-90]{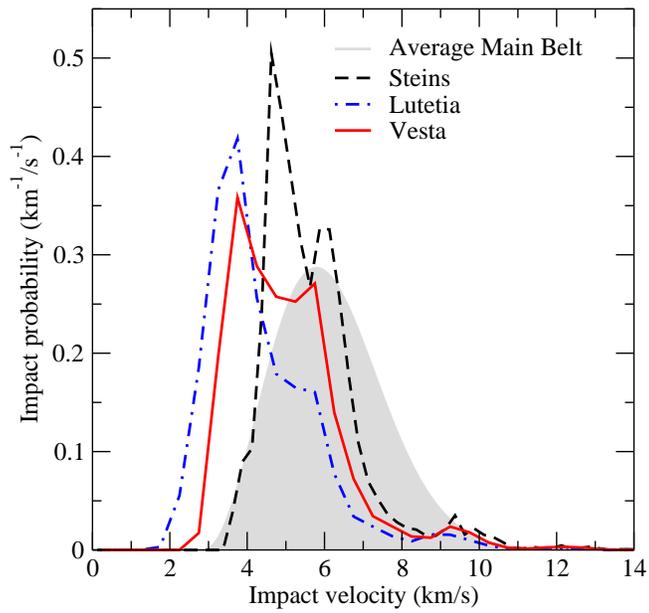}
\caption{Impact velocity distribution for Vesta. For comparison,
  asteroids Steins and Lutetia are also shown, as well as the average
  impact velocity distribution for the main belt. The computed average
  impact velocity for Vesta is 5.0~km/s. { The impact velocities
    are derived using the \cite{far92} algorithm, which computes the
    average impact velocity between asteroids with absolute magnitude $<$13
    and Vesta.}}
\label{vel}
\end{figure*}

In analogy with previous work, the impactor flux is characterized by
its size-frequency distribution and impact velocity distribution.  The
impactor SFD is taken from the model of the main belt population of
\cite{bot05a}.  In this work, we will also consider a second MBA
population derived by a recent survey of small main belt asteroids
\citep{gla09}.  { The latter distribution is valid down to absolute
  magnitudes of $\sim$18, (corresponding to a diameter of about 0.8~km
  for an albedo of 0.2), therefore well above the range of interest
  for this work. We extended it towards smaller sizes ($\sim$10~m) by
  linearly extrapolating the cumulative slope of -1.5 measured in the
  range $0.8-3$~km. We are aware that such distribution does not
  necessarily correspond to the real MBA SFD, however it provides a
  good term of comparison.}  Following the approach of \cite{mar10},
we derive the intrinsic collision probability between MBAs and Vesta
$P_i=2.85\cdot10^{-18}$~km$^{-2}$yr$^{-1}$, as well as Vesta's impact
velocity distribution (see Fig.~\ref{vel}).

For the crater scaling law, we adopted the Pi-group scaling
relationships of \cite{hol07}. These scaling laws allow us to estimate
the { diameter of a crater given the velocity ($v$), diameter
  ($d$)}, and density ($\delta$) of the impactor along with the
density ($\rho$) and strength ($Y$) of the target. In addition to
these quantities, two parameters ($\nu, \mu$) account for the nature
of the terrains (hard-rock, cohesive soil, porous material).  In this
paper, we investigate both hard-rock and cohesive soils scaling laws,
whose parameters are $\nu=0.4, \mu=0.55$ and $\nu=0.4, \mu=0.41$,
respectively.  We assume $\delta=2.6$~g/cm$^3$ and $Y=2\times
10^8$~dyne/cm$^2$ for hard-rock \citep{mar10}. The bulk silicate
density of Vesta is $\rho\sim3.1$~g/cm$^3$ \citep{rus12}.  An
important aspect of crater formation concerns the transition from {\it
  transient} crater size to {\it final} crater size.  { During the
  formation of terrestrial and lunar simple craters, this transition
  occurs when excavation flow has ceased and material still lining the
  transient crater collapses under the influence of gravity to form a
  breccia lens.  The collapse of that material effectively enlarges
  the diameter of the crater slightly \citep{den77,gri77}.  The
  question is if this process occurs also on low-gravity asteroids.
  Measurements of the depth to diameter ($d/D$) ratio and morphology
  of Vestan craters \citep{vin12} show that they exhibit signs of
  gravity-driven modifications comparable to lunar craters. This
  result suggests that Vestan craters undergo the transient to final
  modification, and therefore we applied this correction in our
  analysis, following the approach described by \cite{mar11}.}  We can
also perform a simple natural experiment to estimate the crater
scaling relationship between small craters and MBAs. Consider that the
shape of the crater SFD on Rheasilvia terrains has an inflection point
near $D \sim 1$~km.  This feature corresponds to a similarly-shaped
inflection point in the near-Earth object SFD near $\sim 0.1$~km. At
these sizes, the near-Earth object SFD closely resembles that of MBAs,
which resupplies the NEO population \citep{bot06}.  Taking the
ratio of these two values, we get a factor of $\sim 10$, the same as
that predicted by our crater scaling law relationship.  Thus, we have
increased confidence that the scaling laws used here are reasonable.

{ Using the impactor SFD, the intrinsic collisional probability and
  the crater scaling law described above, we derive the Model
  Production Function for 1 year, {\rm MPF}(1{\rm yr}).} Then, we can
compute absolute surface ages provided we understand both the time
dependence of the impactor flux in the past and how the SFD has varied
over the same time.  While neither component is known a priori,
modeling work suggests the MBA impact rate and their SFD has been
fairly constant, say within a factor of $\sim 2$, for the last
3-3.5~Gyrs \citep{bot05a,bot05b}. The production rate of
kilometer-sized and smaller craters on the Moon, which were caused by
impactors derived from the MBA population, also appears to have been
constant over this time range \citep{neu94}.
{ Also, the match between the crater SFDs on ancient lunar terrains
and the current MBA SFD, suggests the latter has remained unchanged over
the past $\sim4$~Gyr \citep{str05}, although this conclusion applies to
larger impactors ($>$ 0.5~km).}

Assuming all craters that formed on Vesta's surface are retained
(i.e., crater obliteration processes are negligible), the crater MPF
for Vesta at a time $t$ is given by:

\begin{equation}
{\rm MPF}(t)={\rm MPF}(1{\rm yr})\cdot t \label{mpf_eq}
\end{equation}

\noindent where $t$ is the age ($t=0$ is the present). Note that
Equation~\ref{mpf_eq} assumes a constant flux, which is a valid
assumption for the young terrains under study in this
work. The MPF($t$) is used to derive the model cratering age via a
best fit to the observed crater SFD that minimizes the reduced chi
squared value, $\chi_r^2$.  Data points are weighted according to
their Poisson statistics errors.  The formal errors on the best age
correspond to a 50\% increase of the $\chi_r^2$ around the minimum
value. Other sources of uncertainties are neglected \citep[see][for
  more details]{mar11}.

\section{MPF Fitting of the Crater SFDs}

\begin{figure*}[h]
\includegraphics[width=15cm]{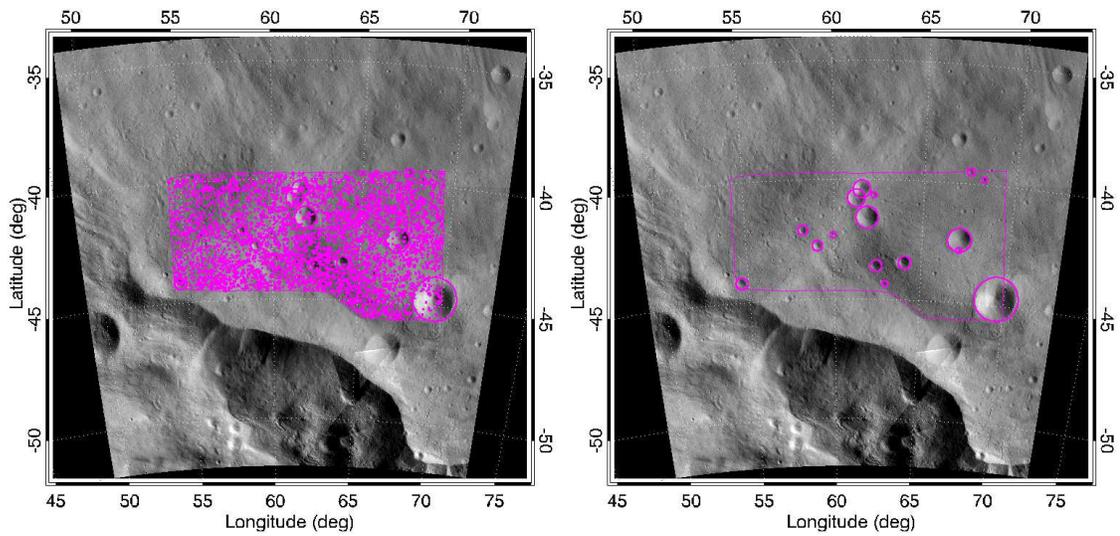}
\caption{Orthographic projection of a LAMO mosaic showing the
  boundaries of the counting region (1200.1~km$^2$) and the craters
  that have been mapped on the Rheasilvia smooth ejecta unit. Left
  panel: 3708 craters larger than 80~m.  Right panel: Crater larger
  that 1~km.}
\label{rs_lamo}
\end{figure*}

\begin{figure*}[h]
\includegraphics[width=12cm,angle=-90]{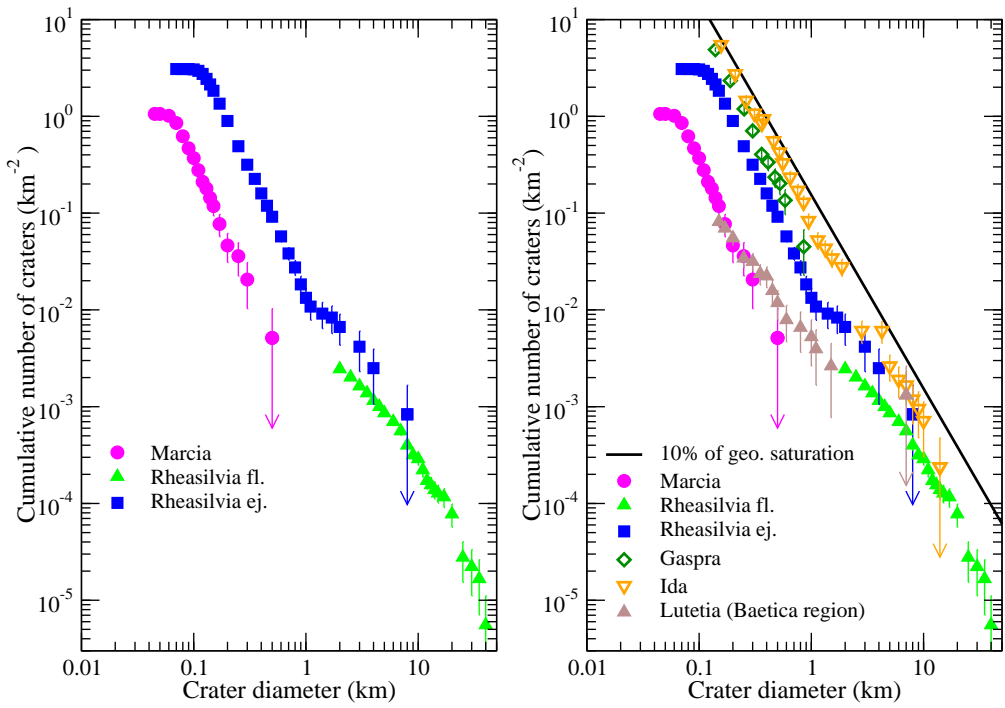}
\caption{Left panel: Crater size-frequency distributions measured on
  Marcia smooth units and Rheasilvia ejecta blanket. Righ panel:
  Comparison with Gaspra and Lutetia. { Green triangles are counts from
  the floor of Rheasilvia \citep{mar12b}.}}
\label{craterSFDs}
\end{figure*}

Figure~\ref{craterSFDs} (left panel) shows the crater SFDs from
Figs.~\ref{marcia_lamo} and \ref{rs_lamo}, as well as those from the
Rheasilvia floor \citep{mar12b}.  We find that Marcia's smooth
terrains have ten times fewer craters per square kilometer than
Rheasilvia's ejecta terrains, implying a significantly older age of
the latter.  We also find an apparent mismatch in crater spatial
density between $D > 1.5$~km craters on the Rheasilvia's ejecta
terrains and those on its floor.  This may seem odd, given that
surfaces with the same age should have the same spatial density.  The
likely explanation is small number statistics. Indeed, {
  Fig.~\ref{rs_lamo} (right panel) shows that the counting area has an
  excess} of large craters compared to surrounding areas, which would
imply we lack statistically significant results for larger craters.

Figure~\ref{craterSFDs} (right panel) shows how these new data compare
with previous measurements on asteroids Gaspra, Ida, and Lutetia (see
Section~1). Gaspra's sub-kilometer crater SFD has a cumulative slope
which is roughly compatible with that observed on Vesta over the same
size range. This is consistent with previous suggestions that Gaspra's
craters are not saturated and are representative of the production
population \citep{cha96a}. It would also argue against the idea that
its observed small craters have been strongly affected by seismic
shaking given that this effect, if present, should affect smaller
bodies to a larger degree than on Vesta \citep[e.g.][]{gre94}.
Moreover, the spatial density of Gaspra's craters is higher than those
on Rheasilvia. This translates into an age of $\sim 1.6$~Gyr, assuming
the same crater scaling law used for Vesta applies to Gaspra. Note
that for Vesta we applied a correction that transforms the transient
crater size into a larger final crater size. It is not clear if such a
correction should apply to Gaspra given its much smaller gravity. On
the other hand, crater $d/D$ ratios on Gaspra \citep{car94} are close
to those found for lunar craters. It is also possible that other
processes, like seismic shaking, may be more effective in producing
shallow craters than gravitational collapse. If we neglect this
correction, the age of Gaspra would become $\sim 3$~Gyr. Gaspra is
often assumed to be a representative member of the Flora family, which
is thought to produce many NEAs \citep{ver08}. Accordingly, this
result may have key implications for the age and evolution of the
Flora family.

Interestingly, our age estimate for Gaspra is significantly older than
previous estimates: $\sim 50$~Myr by \cite{gre94}, $\sim 200$~Myr by
\cite{cha96a}, and $65-100$~Myr by \cite{obr06}. This difference can be
understood in the light of our different assumptions. Age estimates of
Gaspra require one to determine (i) the collision probability of
background MBAs with Gaspra, (ii) the MBA SFD, and (iii) the crater
scaling relationship that can transform projectiles into craters
slamming into Gaspra.  While the methods to achieve (i) have been
known for some time \citep[e.g.][]{opi51,wet67,far92,bot94}, the
assumptions used for (ii) and particularly (iii) by previous works
{ differ} from those made here. For example, Greenberg et al. and
O'Brien et al. both used a crater scaling law based on early hydrocode
simulations \citep{nol96} that, in some cases, assumed that the ratio
between crater to projectile diameters was 20-30, as large or larger
than the expected value for asteroids making craters on the Moon at
$\sim 20$~km/s. More modern estimates, like those used in this paper
\citep[Holsapple and Housen 2007; see discussion in][]{mar10},
however, suggest this value may only be $\sim 10$. This lower value is
also consistent with an interpretation of crater SFDs from Ida and
Eros \citep{botcha06}, as well as our natural experiment
described above. When our code adopts the O'Brien et al. scaling law,
we get an age of $\sim 80$~Myr for Gaspra, very close to their
estimates.

The Vesta and Gaspra distributions both have a lower crater density
than that observed on Ida, which appears to be near
saturation. This would explain why its crater SFD has a shallower
slope than the others.  A lower limit on the age of Ida would be
$\sim3$~Gyr (obtained without considering crater obliteration
processes, thus the real age is probably older).  This age is
consistent with dynamical studies, which suggest the approximate time
needed for Ida and other comparable-sized Koronis family members to
obtain their so-called “Slivan state” spin vectors via the YORP effect
is 2-3~Gyr \citep{vok03}.

Curiously, the crater SFD measured on Lutetia's young Baetica region
is not near saturation, yet it shows a shallower slope than that
observed on Rheasilvia ejecta terrains.  \cite{mar11} suggested this
could be because downslope movement of small debris may have buried
some small craters; there is observational evidence for landslides in
the Baetica region, and its surface has an average gravitational slope
of roughly $25^\circ$.
 
\begin{figure*}[h]
\includegraphics[width=13cm,angle=-90]{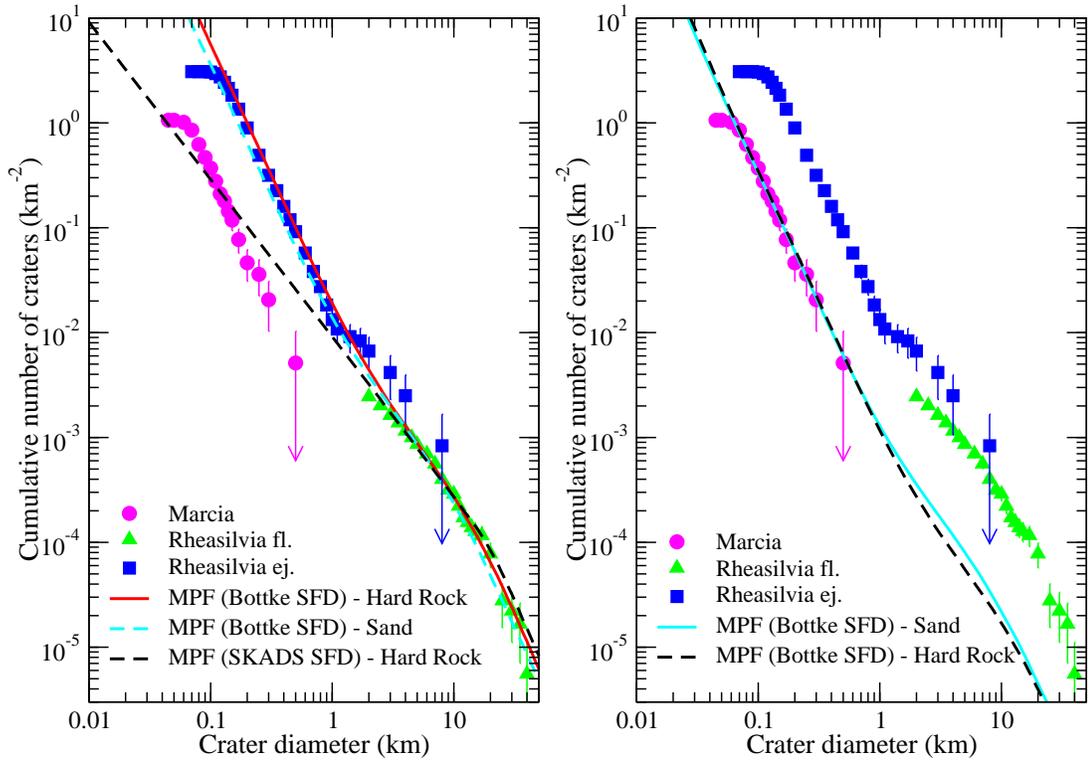}
\caption{MPF-based best fits of the crater SFDS of Rheasilvia (left
  panel) and Marcia (right panel). The MPFs are derived using
  \cite{bot05a,bot05b} MBA SFD and crater scaling laws as
  indicated (see main text for further details). The left-hand panel
  also reports a MPF derived extrapolating the \cite{gla09} MBA SFD
  (valid down to $\sim 0.8$~km projectile size or $\sim 5$~km crater
  size).  The latter is clearly extrapolated beyond the range of
  validity and it is shown only as a comparison (see text).}
\label{mpf}
\end{figure*}

Figure~\ref{mpf} shows the MPF-based best fit to both Rheasilvia
ejecta and Marcia smooth unit crater SFDs.  The red dashed line
indicates the best fit of Rhesilvia floor crater SFD published in
\cite{mar12b}, which was calculated using the hard rock scaling law
\citep{hol07} and a model MBA SFD from \cite{bot05a,bot05b}.
Intriguingly, the model provides a very good fit to both sets of
crater data. This implies that the MBA SFD used to derive the MPF
\citep[namely,][]{bot05a,bot05b} works fairly well down to crater
sizes of about 100~m and projectile sizes of about 10~m.  It is also
consistent with the shape of the MBA SFD remaining roughly constant
over timescales of the order of 0.1~Gyr, 1~Gyr, and 2~Gyr, which
agrees with model predictions from \cite{bot05a,bot05b} { and lunar
  data \citep{str05}}.  We also examined how our results were affected
by the adopted crater scaling law.  We found that computing an MPF
using the crater scaling law for sand produced essentially the same
results, in terms of quality of the fit. However, the actual value of
the material strength has important implication for the age
determination (see below).
 
Another interesting result is that we find most of the Rheasilvia
floor and ejecta crater SFDs fit on the same MPF, which strongly
suggests they were created at the same time that the Rheasilvia
basin-formation event took place.  Interestingly, both Rheasilvia
floor and ejecta contain crater clusters, which, given the lack of
nearby large craters, could be due to self-secondary cratering. Our
results, however, rule out a significant contribution of
self-secondary craters to the populations of craters $> 100$~m.

Similar conclusions can be drawn for the Marcia crater.  First of all,
our MPF does a good job in fitting the observed crater SFD.  If we
apply the same crater scaling law used for Rheasilvia { (i.e. hard
  rock)}, we derive an age of $\sim 60$~Myr. For a comparison, using
the cohesive soil scaling law (and leaving the other parameters
unchanged), the age becomes $\sim 170$~Myr.  This is certainly an
upper limit because it is derived using hard rock strength of
$2\times10^8$~dyne/cm$^2$. Indeed, in the case of cohesive soils, a
reduced strength should be considered. Interestingly, using a more
realistic strength of about a factor of 10 lower would give an age of
$\sim 50$~Myr, which is in agreement with the first
estimate. Therefore it is likely, on the basis of our current data,
that Marcia crater formed about $60$~Myr ago.

\section{Discussions and Conclusions}

Our results have interesting implications for main belt evolution.
First, the crater SFDs on the Marcia and Rheasilvia regions, as well
as the asteroid Gaspra, have approximately the same shape between $0.1
< D < 1$~km.  Given that their crater spatial densities have a wide
range of values, this implies the main belt SFD for asteroid diameters
between $0.01 < D < 0.1$~km has had approximately the same shape for
the last { few Gyr. This is consistent with observations of larger
  impactors from lunar cratering \citep{str05}.}

This outcome matches predictions made by main belt collisional
evolution models \citep[e.g.][]{bot05a,bot05b,obr06}. In these models,
asteroid populations beat up on themselves, with catastrophic
disruption events continually demolishing older asteroids and creating
new fragments.  In these codes, model objects can also be delivered
into the terrestrial planet region by algorithms that try to account
for the fact that small MBAs have their orbits modified by Yarkovsky
thermal drift forces and resonances.  Overall, the model runs of
Bottke et al. and O'Brien et al. show that the MBA SFD quickly
develops a wavy shape and maintains this shape for billions of years.
The pattern only deviates from this in runs for the occasional case
where a very large main belt object { is disrupted} (i.e.,
something hundreds of kilometers in diameter).  Even then, the
characteristic SFD quickly returns { to the original equilibrium
  shape.}

An interesting question is whether the absolute number of small MBAs
has also remained more or less constant over billions of years. This
would imply MBAs were in a quasi-steady state over the same time
period.  Note that such behavior is a common feature of the best fit
runs of \cite{bot05a,bot05b}. Although the absolute ages of
cratered terrains on Vesta and Gaspra are unknown, we can glean
insights into this issue by examining the impact history of the Moon.

Dynamical studies and interpretations of the cratered SFDs on lunar
terrains suggest the main asteroid belt has been the dominant source
of lunar impactors for the { last $\sim$4~Gyr
  \citep{bot02,iva02,kri02,mor03,str05}}.  Several terrains on the
lunar nearside have absolute ages that were determined directly (and
indirectly) using samples returned by the Apollo and Luna missions.
Combining these values with measurements of the spatial density of
small craters on these surfaces, it is possible to estimate the
average lunar impact { flux at several moments in time between the
  present day and $\sim$3.5~Gyr ago} \citep[see][for a recent review of
  these issues, as well as Neukum and Ivanov 1994]{sto01}.  The most
straightforward interpretation of these values is that the average
lunar impact flux for { MBA-derived projectiles} has remained
approximately constant (i.e., within a factor of 2) over the last {
  few Gyr} \citep{har81,mce97,sto01,mar09,hie12}.

Accordingly, if the lunar impact flux has not changed very much over a
{ $\sim$3~Gyr interval}, and nothing else important has changed
regarding the delivery of small MBAs to the terrestrial planet region,
it is logical to think that the small MBA population has remained more
or less constant over this time period as well.  Accordingly, we
predict the small MBA population, on average has been near or in
steady state for { several Gyr}.

Note that the match obtained between model results and observations
probably means the assumptions made by the collision codes are broadly
reasonable (e.g., collision probabilities, disruption scaling laws,
nature of their fragment SFDs, dynamical depletion rates, etc.).  With
this said, one must be careful in how this interpretation is used;
while the results of the best fit collision models indeed match
numerous constraints, they may not yet provide unique solutions.

{ For instance,} small asteroids can also lose material via the
non-gravitational YORP spin-up mechanism \cite[e.g.][]{pra10}.
Interestingly, the older collisional codes discussed above do not
include this effect, yet they appear to reproduce the small MBA SFD
and many other constraints. This could suggest that YORP mass
shedding, as a process, is secondary to other mass loss mechanisms,
and that it is reasonable to neglect it when modeling the collisional
evolution of the small MBA SFD. It is also possible, however, that
other aspects of the { modeling} are incorrect and are in effect
compensating for the absence of YORP mass shedding.  If so, a closer
analysis of this mass shedding process, combined with our new crater
constraints, may yield new insights into the main belt evolution
\citep{ros12}.  We consider this an interesting issue for future work.

The apparent stability of the small MBA SFD in terms of both shape and
population over time is perhaps surprising.  It indicates that the
sources of small MBAs, such as asteroid family forming events, are
well balanced by the various sink mechanisms. Modeling work suggests
that large asteroid disruption events are capable of supplying a
sizeable fraction of the total MBA population for hundreds of Myr to
Gyr timescales via a collisonal cascade \citep{bot05a,bot05b}.
Thus, at any given time, the small MBA population may be dominated by
fragments from several tens of large breakup events. This likely
explains why the samples in our meteorite collections only appear to
represent about 30 parent bodies \citep{mei99,kei00,kei02,bur02,bot05c}.

In summary, we have presented in the paper the sub-kilometer crater
SFDs of two young terrains on Vesta. We found that these crater SFDs
match the estimated crater production function derived from model main
belt SFDs. We used the { modeled} crater production function to
estimate the ages of our two young regions.  We find an age of
$\sim$1~Gyr for Rheasilvia basin and $\sim$60~Myr for Marcia crater.\\

{\bf Acknowledgments} We thank Nadine G.~Barlow and an anonymous
referee for valuable comments that improved the manuscript.
D.P.~O'Brien thanks NASA's Dawn at Vesta Participating Scientist
Program. The contributions of Simone~Marchi, William F.~Bottke and
David A.~Kring were supported by the NASA Lunar Science Institute
(Center for Lunar Origin and Evolution at the Southwest Research
Institute in Boulder, Colorado NASA Grant NNA09DB32A; Center for Lunar
Science and Exploration at the Lunar and Planetary Institute in
Houston, Texas).


\newpage

\end{document}